\documentclass[a4paper,12pt]{article}
\usepackage{amsthm,amssymb,amsmath,url,amsfonts,mathrsfs,bm,}
\usepackage{graphicx}% %\usepackage{amsfonts,amsmath,amssymb,mathrsfs,array}
\theoremstyle{definition}

\newcommand{\bal}{\begin{align*}}
\newcommand{\bea}{\begin{eqnarray*}}
\newcommand{\eea}{\end{eqnarray*}}
\newcommand{\el}{{(l)}}

\newcommand{\elm}{{(l-1)}}
\newcommand{\en}{{(n)}}
\newcommand{\enp}{{(n+1)}}

\newcommand{\enm}{{(n-1)}}
\renewcommand{\em}{{(m)}}
\newcommand{\st}{{(l,m,n)}}

\newcommand{\emm}{{(m-1)}}

\newcommand{\cm}{{\text{cm}}}

\newcommand{\cz}{{\bar z}}

\newcommand{\eal}{\end{align*}}

\newcommand{\tit}{\textit}
\newcommand{\La}{\Lambda}

\newcommand{\ity}{\infty}

\newcommand{\no}{{(1)}}

\newcommand{\Z}{\textbf{Z}}

\newcommand{\mo}{{-1}}

\newcommand{\bg}{\begin}

\newcommand{\tx}{\text}
\newcommand{\sra}{\stackrel{\leftrightarrow}}
\newcommand{\sraa}{\stackrel{\longleftrightarrow}}
\newcommand{\pa}{\partial}

\newcommand{\na}{\nabla}

\newcommand{\De}{\Delta}

\newcommand{\sm}{\setminus}

\newcommand{\lam}{\lambda}

\newcommand{\HH}{{\cal H}}

\newcommand{\bo}{{\bm \omega}}

\newcommand{\om}{\omega}

\newcommand{\C}{\mathbb{C}}
\newcommand{\R}{\mathbb{R}}

\newcommand{\DD}{{\cal D}}
\newcommand{\bz}{{\bar{\mathbf{z}}}}
\newcommand{\Si}{\Sigma}

\newcommand{\ep}{{(p)}}

\newcommand{\la}{\langle}

\newcommand{\ra}{\rangle}
\newcommand{\x}{{\bf x}}

\newcommand{\y}{{\bf y}}

\newcommand{\z}{{\bf z}}

\newcommand{\bn}{{\bm \nabla}}

\usepackage{graphicx}% Include figure files

\begin{document}
\begin{large}
\title{\bf {Quantum field theory without divergence: the method of the interaction operators}}
\end{large}
\author{Bruno Galvan \footnote{e-mail: b.galvan@virgilio.it}\\ \small via Melta 16, 38121 Trento, Italy.}
%\date{\small September 2007}
\maketitle
\bg{abstract}
The recently proposed interior boundary conditions approach [S. Teufel and R. Tumulka: Avoiding Ultraviolet Divergence by Means of Interior Boundary Conditions, arXiv:1506.00497] is a method for defining Hamiltonians without UV divergence for quantum field theories. In this approach the interactions between sectors of the Fock space with different number of particles (inter-sector interactions) are obtained by extending the domain of the free Hamiltonian to include functions with singularities. In this paper a similar but alternative strategy is proposed, in which the inter-sector interactions are implemented by specific interaction operators. In its simplest form, an interaction operator is obtained by symmetrizing the asymmetric operator $\|\hat{\textbf{x}} \|^{-1} \Delta \|\hat {\textbf{x}}\|^{-1}$. The inter-sector interactions derive from the singularities generated by the factors $\|\hat{\textbf{x}}\|^{-1}$ enclosing the Laplacian, while the domain of the interaction operator does not include singular functions. As a consequence the interaction operators and the free Hamiltonian have a common dense domain, and they can be added together to form the complete Hamiltonian with interaction.
\end{abstract}

\section{Introduction}

The Hamiltonians of most quantum field theories (or, more in general, of quantum theories of systems with variable number of particles) are ill defined because their interaction terms contain unbalanced creation operators (see for example \cite{g}). The consequence is that these Hamiltonians are plagued by ultraviolet divergence.

Standard methods for avoiding this problem include discretizing space or smearing the creation operators over a small region of space. A non-standard method for solving this problem has been proposed by the author some years ago \cite{g}, but it has revealed itself to be not fruitful.

Recently a promising approach has been proposed by Teufel et al. \cite{t1,t2}, the \tit{interior boundary conditions} (IBC) approach. In this approach the domain of the free Hamiltonian is enlarged with the inclusion of singular functions. Due to these singularities the free Hamiltonian is no longer self-adjoint. For restoring self-adjointness it is necessary adding a term to the Hamiltonian connecting different sectors of the Fock space, and imposing suitable inter-sector boundary conditions to the domain of definition. The result is that the probability is no longer conserved within the single sectors, and inter-sector interactions take place\footnote{Actually the basic idea of this approach had been previously considered by other authors \cite{thomas,yafev}, and in particular by Moshinsky \cite{mo1,mo2,mo3}, who also developed a relativistic version of this approach \cite{mo3}. However Teufel et al. have explicitly applied this idea for the first time to the development of a realistic quantum field theory without ultraviolet divergence.}. In their papers Teufel et al. have announced some interesting results that suggest that the IBC approach is physically relevant and not merely a mathematical curiosity.

In this paper a different approach is proposed, which however has some similarities with the IBC approach. In the novel approach the inter-sector interactions are implemented by specific interactions operators which are added to the free Hamiltonian. In the simplest case, an interaction operator is obtained by symmetrizing the asymmetric operator
\begin{equation}
\frac{1}{\|\hat \x\|} \De \frac{1}{\|\hat \x\|}.
\end{equation}
The above operator has a domain without singular functions, and it is non-symmetric because of the singularities generated by the factors $\|\hat \x\|^\mo$ enclosing the Laplacian. Also in this case the above operator is symmetrized by adding an inter-sector term and by imposing suitable inter-sector boundary conditions.

The interaction operators and the free Hamiltonian (i.e. the normal Laplacian in the non-relativistic case) have (arguably) a common dense domain, and therefore they can be added together to form the complete Hamiltonian with interaction.

After in introductory pedagogical example, the interaction operators are developed in this paper for a system composed of three types of non-relativistic spinless particles, namely ``electrons'', ``antielectrons'' (collectively called fermions), and bosons. These particles interact through two distinct types of processes: (1) the bosons are emitted/absorbed by the fermions and (2) pairs of electrons-antielectrons are created/annihilated from/in a boson. 

This paper is only a preliminary proposal, and only the definition of the interacting operators and some of their elementary properties are presented. Further studies are necessary in order to verify if this approach is mathematically consistent and physically relevant.

%*****************************************************************************************************************

\section{Introductive example: a fixed source emits/absorbs a single boson} \label{se1}

In this simple pedagogical example a source at the point $\x=0$ emits and absorbs a single boson. The Hilbert space is $\HH:=L^2(\R^3) \oplus \C$, and the vectors of $\HH$ are of the form $(\psi(\x), c)$.

The asymmetric interaction operator $\La$ on $L^2(\R^3)$ is defined as follows:
\bg{equation} \label{ppu}
\La:= \frac{R}{\|\x\|}\De \frac{R}{\|\x\|},
\end{equation}
where 
\begin{equation}
[R\psi](\x):=\frac{1}{4 \pi}\int_{S^2} \psi(\|\x\|\bo) d^2 \om = \frac{1}{\sqrt{4\pi}}\psi_{00}(\|\x\|)
\end{equation}
is the projector onto the subspace of $L^2(\R^3)$ composed of the wave functions with radial symmetry. In the above expression $\psi_{00}(r)$ is the coefficient with $\ell=m=0$ of the spherical Harmonic expansion of $\psi$; the generic coefficient has the form $\psi_{\ell m}(r)$.

A simple calculation shows that in spherical coordinates $\La$ reads
\begin{equation}
\La = \frac{1}{r^2} \pa^2_r.
\end{equation}

A suitable domain for $\La$ is the domain $\DD^\no$ composed of the wave functions $\psi \in L^2(\R^3)$ which: (i) have compact support, (ii) are continuous everywhere, (iii) are smooth in $\R^3 \sm \{0\}$, and (iv) satisfy $\|\La \psi\| < \ity$. Since $\psi$ is not smooth in $0$ it may be that $\psi'_{00}(0) \neq 0$, where the apex denotes the derivative by $r$. Note moreover that $\psi(0)=\psi_{00}(0)/\sqrt{4\pi}$. In Fig. 1 a typical element of $D^\no$ is schematized.
\vspace{0mm}
\begin{center}
%\begin{figure}
\includegraphics {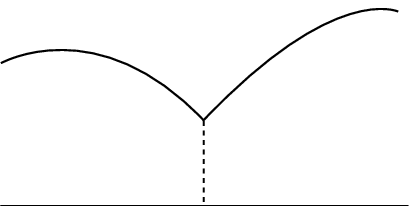} \\
%\end{figure}
\small{Fig. 1}
\end{center}
 
The operator $\La$ is not symmetric on $\DD^\no$, but we have:
\begin{equation} \label{a1}
 \la \phi |\sra{\La}|\psi \ra := \la \phi |\La \psi \ra - \la \La \phi |\psi \ra = \phi'^*_{00}(0) \psi_{00}(0) - \phi^*_{00}(0) \psi'_{00}(0).
\end{equation}
In fact:
\bea
& & \la \phi |\La \psi \ra =  \int \psi^* \frac{1}{r^2} \pa^2_r R \psi \, r^2 dr d^2 \bo = \int \phi_{00}^* \psi''_{00} dr= \\
& & \phi_{00}^* \psi'_{00} \Big |_0^\ity - \int \phi^{'*}_{00} \psi'_{00} dr = 
-\phi^*_{00}(0)\psi'_{00}(0) - \int \phi^{'*}_{00} \psi'_{00} dr ,
\eea
from which equation (\ref{a1}) easily follows.

Let us define the operators $B, C:D^\no \to \C$ as follows:
\bg{eqnarray} 
& & B \psi := \psi_{00}(0)/\sqrt{4\pi} = \psi(0), \label{daa}\\
& & C\psi := \sqrt{4\pi} \, \psi'_{00}(0) = \left. \frac{d}{dr}  \int_{S^2} \psi(r \bo) d^2 \bo \right|_{r=0}. \label{dbb}
\end{eqnarray}
By using the operators $B$ and $C$, equation (\ref{a1}) reads:
\begin{equation} \label{a3}
\la \phi |\sra{\La}|\psi \ra = \la C \phi|B \psi \ra- \la B \phi|C \psi\ra.
\end{equation}

In order to obtain the symmetric interaction operator $\La_s$ on $\HH$ let us extend first the action of $\La$ and $C$ from $L^2(\R)$ to $\HH$ by defining $\La c = C c = 0$. The operator $\La_s$ is then defined as follows:
\begin{equation}
\La_s := \La + C,
\end{equation}
and it is symmetric on the domain 
\begin{equation}
\DD_\La:= \{(\psi, c) \in  \HH: \psi \in \DD^\no \tx{ and } c = B\psi \},
\end{equation}
which is dense in $\HH$. Note that the elements of $\DD_\La$ satisfy the boundary condition
\bg{equation} \label{dbc}
\psi(0)=c.
\end{equation}
This boundary condition and the others that will be introduced in the next sections are so natural that can be qualified as \tit{standard}.

Let us prove that $\La_s$ is symmetric and negative on $\DD$. Symmetry:
\[
\la (\phi, B \phi) |\sra{C}|(\psi, B \psi) \ra = \la B \phi |C \psi \ra - \la C \phi |B \psi \ra,
\]
which together with equation (\ref{a3}) proves the thesis. Negativity:
\bg{eqnarray*}
& & \la (\psi, B \psi) |\La_s (\psi, B \psi) \ra =  \la \psi| \La \psi \ra  + \la B \psi |C \psi \ra  =  \int \psi_{00}^* \psi^{''}_{00} dr + \la B\psi|C \psi \ra = \label{poi} \\
& & = \psi_{00}^* \psi'_{00} \Big |_0^\ity - \int |\psi'_{00}|^2 dr + \la B\psi|C \psi \ra \leq 
 - \psi_{00}^*(0) \psi'_{00}(0) + \la B\psi|C \psi \ra = 0. \nonumber
\end{eqnarray*}

The proposed symmetrization of $\La$ is not the only possible, as one can easily see by simply replacing the above definitions of the operators $B$ and $C$ with the followings: $B\psi:=\psi_{00}(0)/\lam$ and $C\psi:=\lam \psi'_{00}(0)$, where $\lam \in \R \sm \{0\}$. However in this case the standard boundary condition (\ref{dbc}) is no longer satisfied. If instead the operator $\lam B$ is added to $\La_s$ the standard boundary condition is satisfied and one still obtains a symmetric interaction operator (note that $B$ is symmetric on $\DD_\La$).

Let us return to the main subject. The normal Laplacian, which is usually proportional to the free Hamiltonian, is well defined, symmetric and negative on $\DD^\no$. As a consequence, the free Hamiltonian and an interaction term proportional to $\La_s$ can be added together to form the complete Hamiltonian with interaction. For example we can define:
\begin{equation} \label{pori}
H_0 (\psi, c) := \left (- \frac{\De}{2m} \psi + E \psi, E \right), \; \;  \; H_I := - g \La_s.
\end{equation}
where $E \geq 0$ is the energy of the source and $g>0$ is a suitable coupling constant. The complete Hamiltonian is therefore:
\begin{equation} \label{ham}
H = H_0 + H_I,
\end{equation}
which is symmetric and positive on the dense domain $\DD_\La$. A theorem of functional analysis states that a semi-bounded symmetric operator admits a distinguished self-adjoint extension, the so called Friedrich extension \cite[p. 176]{reed}. Further studies are necessary to verify if $H$ is also essentially self-adjoint on $\DD_\La$.

Let us conclude with a remark relative to the presence of the operator $R$ in the definition (\ref{ppu}). One can easily see that $\psi_{\ell m}(0)= 0$ for $\ell >0$ and $\psi \in \DD^\no$. As a consequence the inter-sector interaction (the asymmetry of $\La$) is only determined by the component of the wave function with null angular momentum. The operator $R$ limits therefore the action of the Laplacian of the interaction operator to the minimum necessary for determining inter-sector interactions.

%*****************************************************************************
\section{The general case}

As previously said, the general system we study is composed of three types of non-relativistic spinless particles, namely: ``electrons'' $e$, ``antielectrons'' $\bar e$ and bosons $b$. Electrons and antielectrons are collectively called fermions. Two types of interaction processes are considered, namely: (1) the bosons are emitted/absorbed by the fermions and (2) pairs of electrons-antielecrons are created from a boson or annihilate in a boson. These two processes will be implements by two different interaction operators. Obviously this model mimics electrodynamics, where the bosons correspond to the photons.

The Hilbert space of this the system is 
\begin{equation} \label{ds}
\bigoplus_{l,m,n=0}^\ity \HH^\el_e \otimes \HH^\em_{\bar e} \otimes \HH^\en_b,
\end{equation}
In the above expression $\HH^\el_e$ and $\HH^\em_{\bar e}$ are the antisymmetric subspaces of $L^2(R^3)^{\otimes \, l}$ and $L^2(R^3)^{\otimes m}$, respectively, and $\HH^\en_b$ is the symmetric subspace of $L^2(\R)^{\otimes n}$. The variables of the electrons, antielectrons, and bosons are $z^\el = (\z_1, \ldots, \z_l)$, $\cz^\em = (\bz_1, \ldots, \bz_m)$ and $x^\en = (\z_1, \ldots, \z_n)$, respectively. The addends of the direct sum (\ref{ds}) are called the \tit{sectors} of $\HH$, and the following notation is adopted:
\begin{equation}
\HH^{(l,m,n)}:= \HH^\el_e \otimes \HH^\em_{\bar e} \otimes \HH^\en_b.
\end{equation}
Often also the following ``multi-sector'' subspace
\begin{equation}
\HH^{(l,m)}:= \HH^\el_e \otimes \HH^\em_{\bar e} \otimes \left (\oplus_{n=0}^\ity \HH^\en_b \right ).
\end{equation} 
will be utilized. The symbol $\psi^{(l,m,n)}$ ($\psi^{(l,m)}$) denotes the projection of $\psi \in \HH$ onto the subspace $\HH^{(l,m,n)}$ ($\HH^{(l,m)}$). If $X$ is an operator on $\HH$, the symbol $X^{(l,m,n)}$ ($X^{(l,m)}$) denotes its restriction to the subspace $\HH^{(l,m,n)}$ ($\HH^{(l,m)}$).

It is useful to characterize two type of operators: (i) the \tit{intra-sector} operators, for which $X \psi^\st \in \HH^\st$, and (ii) the \tit{inter-sector} operators, for which $X \psi^\st \in \HH^{(l',m',n')}  \ne \HH^{(l,m,n)}$. 

The common dense domain $\DD$ of all the operators of the theory, generically represented by $X$, is defined as follows. For any sector $\st$ the domain $\DD^\st$ is composed of the vectors $\psi^\st$ that: (i) have compact support, (ii) are continuous everywhere, (iii) are smooth in $\R^{3(l+m+n)} \sm C^{(l,m,n)}$ (see below), and (iv) satisfy $\|X \psi^\st\| < \ity$ for all $X$. The set $C^{(l,m,n)}$ mentioned in the point (iii) is the collision set of the interacting particles, namely:
\begin{equation}
C^{(l,m,n)}:=\{(z^\el; \cz^\em; x^\en): \x_i=\z_k \tx{ or } \x_i = \bz_k \tx{ or } \z_h = \bz_k \tx{ for some } i,h,k\}.
\end{equation}
The domain $\DD$ is then the subspace of $\bigoplus_\st \DD^\st$ composed of the vectors $\psi$ such that $\sum_\st \|X \psi^\st\|^2 < \ity$ for all $X$.

The normal Laplacian is assumed to be well defined and symmetric on $\DD$. Two other important domains are obtained by imposing suitable boundary conditions to the vectors of $\DD$, as explained in the next paragraph.

The two interacting operators implementing the emission/absorption process and the creation/annihilation process will be denoted by $\La_s$ and $\Si_s$, respectively, and will be defined in the next two subsections. The method for defining the two operators is similar, and for $\La_s$ it is the following: a non-symmetric operator $\La$, which is the generalization of the operator $\La$ introduced in the previous section, is defined on $\DD$. This operator it then symmetrized by adding a suitable inter-sector operator $C_\La$ to it, and by imposing suitable inter-sector boundary conditions to the vectors of $\DD$, obtaining in this way the domain $\DD_\La$. The operator $\La_s:= \La + C_\La$ is therefore symmetric on $\DD_\La$. Analogously, the operator $\Si_s:= \Si + C_\Si$ is symmetric on $\DD_\Si$. There is however also an important difference in the definition of the two operators, that will be emphasized in the last section.

A reasonable complete Hamiltonian with interaction for our system is therefore 
\begin{equation}
H := H_0 - h \La_s - g\Si_s,
\end{equation}
where $H_0$ is the free Hamiltonian based on the normal Laplacian and $h$ and $g$ are two positive coupling constants.

In this paper it has been only proved that $H$ is symmetric on $\DD_H:= \DD_\La \cap \DD_\Si$. Further studies are necessary for (possibly) proving that $\DD_H$ is dense in $\HH$, that $H$ is essentially self-adjoint on $\DD_H$, and of course that $H$ has a spectrum and other properties that are interesting from the physical point of view.

\subsection{The emission/absorption operator}

In order to simplify the notation in this subsection, given the sector $\HH^{(l,m,n)}$ let us define
\begin{equation}
y^\ep = (\y_1, \ldots, \y_p):=(\z_1, \ldots, \z_l; \bz_1, \ldots, \bz_m), \tx{ where } p := l+m.
\end{equation}

The asymmetric interaction operator $\La$ is defined as follows: 
\begin{equation}
\La^{(l,m,n)} := \sum_{i=1}^n \left (\frac{q(1) R^i_1}{\|\hat \x_i - \hat \y_1\|} + \ldots + \frac{q(p) R_p^i}{\|\hat \x_i - \hat \y_p\|} \right ) \De_{\x_i} \left ( \frac{q(1) R_1^i}{\|\hat \x_i - \hat \y_1\|} + \ldots + \frac{q(p) R_p^i}{\|\hat \x_i - \hat \y_p\|} \right ),
\end{equation}
where $q(k)$ is the charge of the $k$-th particle, namely 
\begin{equation}
q(k):=
\bg{cases}
-1 \tx{ if } k \leq l \; \; (\y_k \tx{ is an electron}),\\
+1 \tx{ if } k > l \; \; (\y_k \tx{ is an anti-electron}),
\end{cases}
\end{equation}
and
\begin{equation}
[R_k^i \psi^{(l,m,n)}](y^\ep, x^\en) := 
\frac{1}{4\pi} \int_{S^2} \psi^{(l,m,n)} (y^\ep, \x_1, \ldots, \underbrace{\y_k+ \|\x_i-\y_k\| \bo}_{i\tx{-th x-place}}, \ldots, \x_n)  d^2 \om
\end{equation}
for $1 \leq i \leq n$ and $1 \leq k \leq p$, and $R_k^i\psi^{(l,m,n)}=0$ otherwise. In words, $R_k^i$ is the projector onto the subspace composed of the vectors in which the coordinate $\x_i$ has radial symmetry relative to the center $\y_k$.

It is useful to decompose $\La$ as follows:
\begin{equation}
\La= M + V,
\end{equation}
where 
\begin{equation}
M^{(l,m,n)} := \sum_{k=1}^p \sum_{i=1}^n M^i_k \; \; \tx{ and } \; \; V^{(l,m,n)}:= \sum_{h < k}^p \sum_{i=1}^n V^i_{hk},
\end{equation}
where in turn
\bg{eqnarray}
& M_k^i := & \frac{R_k^i}{\|\hat \x_i - \hat \y_k\|} \De_{\x_i} \frac{R_k^i}{\|\hat \x_i - \hat \y_k\|},\\
& V^i_{hk}:= & q(h)q(k) \left (\frac{R_h^i}{\|\hat \x_i - \hat \y_h\|} \De_{\x_i} \frac{ R_k^i}{\|\hat \x_i - \hat \y_k\|} +
\frac{R_k^i}{\|\hat \x_i - \hat \y_k\|} \De_{\x_i} \frac{R_h^i}{\|\hat \x_i - \hat \y_h\|} \right).
\end{eqnarray}
A reasonable conjecture is that the term $M$ is responsible for the dressing of the fermions by the bosons, and that the terms $V$ determines the potential between the fermions. Note that the operator $V^i_{hk}$ relative to two fermions of the same type has opposite sign with respect to that relative to two fermions of different type. This arguably leads to a repulsive potential for fermions of the same type and to an attractive potential for fermions of different type.

As said before, $\La$ is not symmetric on $\DD$, and in particular one can prove that
\begin{equation} \label{inp}
\la \phi^{(l,m)} |\sra{\La}|\psi^{(l,m)} \ra = \sum_{k=1}^p \la C_k N_b \phi^{(l,m)} |B_k \psi^{(l,m)} \ra - \la B_k \phi^{(l,m)} |C_k N_b \psi^{(l,m)} \ra.
\end{equation}
where $N_b$ is the number operator for the bosons, and $B_k$ and $C_k$ are two inter-sector operators from $(l,m,n)$ to $(l,m, n-1)$ defined as follows:
\bg{eqnarray}
& & [B_k \psi^{(l,m,n)}](y^\ep, x^\enm):=\psi^{(l,m,n)}(y^\ep, \y_k, x^\enm), \\
& & [C_k \psi^{(l,m,n)}](y^\ep, x^\enm):=\left. \pa_r \int_{S^2} \psi^{(l,m,n)}(y^\ep, \y_k+ r \bo , x^\enm) d^2 \om \right |_{r=0}
\end{eqnarray}
for $n \geq 1$ and $1 \leq k \leq p$, and $B_k\psi^{(l,m,n)}=C_k\psi^{(l,m,n)}=0$ otherwise.

In the equation (\ref{inp}) the only contribute to the asymmetry of $\La$ comes from $M$, because one can prove that $V$ is symmetric on $\DD$:
\begin{equation} \label{inp2}
\la \phi |\sra{V}|\psi \ra =0.
\end{equation}
Equations (\ref{inp}) and (\ref{inp2}) are proved in the appendix.

As previously explained, for symmetrizing $\La$ let us define $\DD_\La$ as the subspace of $\DD$ composed of the vectors satisfying the boundary conditions
\begin{equation}
B_k \psi^{(l,m)} = \psi^{(l,m)} \tx{ for } k=1, \ldots, p \tx{ and all } (l,m).
\end{equation}
An equivalent way to express these boundary conditions is:
\begin{equation}
\psi^{(l,m,n)}(y^\ep; \y_k, x^\enm)=\psi^{(l,m,n-1)} (y^\ep; x^\enm) \tx{ for } k=1, \ldots, p \tx{ and } n=1, 2, \ldots .
\end{equation}

Let us define finally the operator
\begin{equation}
C^{(l,m)}_\La = \sum_{k=1}^p C_k N_b.
\end{equation}
The symmetric interacting operator $\La_s$ is then defined as follows:
\begin{equation}
\La_s:= \La + C_\La.
\end{equation}
One can easily prove that $\La_s$ is symmetric on $\DD_\La$. In fact:
\bea
& & \la \phi^{(l,m)} |\sra{S}_\La |\psi^{(l,m)} \ra = \sum_{k=1}^p \la \phi^{(l,m)} |C_k N_b \psi^{(l,m)} \ra - \la C_k N_b \phi^{(l,m)} |\psi^{(l,m)} \ra =\\
& & = \sum_{k=1}^p \la B_k \phi^{(l,m)} |C_k N_b\psi^{(l,m)} \ra - \la C_k N_b\phi^{(l,m)} |B_k \psi^{(l,m)} \ra, 
\eea
from which the thesis easily follows. Since $V$ is symmetric on $\DD$ then $M + C_\La$ is symmetric on $\DD_\La$, and one can also prove that it is negative, that is
\begin{equation}
\la \psi |(M + C_\La) \psi \ra \leq 0 \tx{ for } \psi \in \DD_\La.
\end{equation}
The proof is similar to the proof given in section \ref{se1}, and it is omitted.

%***************************************************************************
\subsection{The creation/annihilation operator}

The asymmetric creation/annihilation operator $\Si$ is defined as follows:
\begin{equation}
\Si^{(l,m)}:= \sum_{h=1}^l \sum_{k=1}^m \Si_{hk},
\end{equation}
where
\begin{equation}
\Si_{hk}:= \frac{R_{hk}}{\|\hat \z_h - \hat \bz_k\|} \frac{(\bn_{\z_h}-\bn_{\bz_k})^2}{4} \frac{R_{hk}}{\|\hat \z_h - \hat \bz_k\|}.
\end{equation}
In the above expression
\begin{equation}
[R_{hk} \psi^{(l,m,n)}](\ldots, \z_h, \ldots, \bz_k, \ldots) = \frac{1}{4 \pi} \int_{S^2} \psi^{(l,m,n)}(\ldots, \underbrace{\Z + \|\z\| \bo /2}_{h\tx{-th } z \tx{ place}}, \ldots, \underbrace{\Z - \|\z\| \bo /2}_{k\tx{-th } \cz \tx{ place}}, \ldots) d\bo
\end{equation}
for $1 \leq h \leq l$ and $1 \leq k \leq m$, and $R_{hk} \, \psi^{(l,m,n)}=0$ otherwise, where
\begin{equation}
\Z:= \frac{\z_h + \bz_k}{2} \tx{ and } \z:= \z_h - \bz_k.
\end{equation}
are the coordinates of the center of mass of the pair $(\z_h, \bz_k)$ (consider that electrons and antielectrons have necessarily the same mass).

In order to better understand the meaning of the above definition, let us introduce the unitary map $U_{hk}: \psi \mapsto \psi_{hk}$, where $\psi_{hk}$ is the wave function $\psi$ expressed in the coordinates of the center of mass of $(\z_h, \bz_k)$, namely:
\begin{equation}
\psi_{hk} (\ldots, \underbrace{\Z}_{h\tx{-th } z \tx{ place}}, \ldots, \underbrace{\z}_{k\tx{-th } \cz \tx{ place}}, \ldots):=
\psi (\ldots, \underbrace{\Z+ \z/2}_{h\tx{-th } z \tx{ place}}, \ldots, \underbrace{\Z-\z/2}_{k\tx{-th } \cz \tx{ place}}, \ldots).
\end{equation}
One can see easily that
\begin{equation}
\frac{(\bn_{\z_h}-\bn_{\bz_k})^2}{4} =  U_{hk}^\mo \De_\z U_{hk} \tx{ and } R_{hk}=U^\mo_{hk} R_\z U_{hk},
\end{equation}
where 
\begin{equation}
[R_\z \psi_{hk}^{(l,m,n)}] ( \ldots, \z, \ldots) := \frac{1}{4 \pi} \int_{S^2} \psi_{hk}^{(l,m,n)}(\ldots, \|\z\| \bo, \ldots) d\bo.
\end{equation}
is the projector onto the subspace of the wave functions of the type $\psi_{hk}$ in which the variable $\z$ has radial symmetry around $0$. As a consequence
\begin{equation}
\Si_{hk} =  U^\mo_{hk} \frac{R_\z} {\|\hat \z\|} \De_\z \frac{R_\z}{\|\hat \z\|} U_{hk}.
\end{equation}
One can therefore see that $R_{hk}$ is the projector onto the subspace of $\HH^{(l,m,n)}$ composed of the functions in which the variables $\z_h$ and $\bz_k$ have radial symmetry around their center of mass, and that in the coordinates of the center of mass of the pair $(\z_h, \bz_k)$ the operator $\Si_{hk}$ has the same form than the operator $\La$ introduced in section \ref{se1}. 

One can prove that 
\begin{equation} \label{poti}
\la \phi |\sra{\Si} |\psi \ra =\la C N_e N_{\bar e} \phi |B \psi \ra - \la B  \phi |C N_e N_{\bar e}\psi \ra, 
\end{equation} 
where $N_e$ and $N_{\bar e}$ are the number operators of the electrons and of the antielectrons, respectively, and the inter-sector operators $B$ and $C$ from $(l,m,n)$ to $(l-1,m-1,n+1)$ are defined as follows:
\bg{eqnarray}
& & [B\psi^{(l,m,n)}](z^\elm, \cz^\emm, x^\enp) = \\
& & = \psi^{(l,m,n)}(\x_1, z^\elm;  \x_1, \cz^\emm; \x_2, \ldots, \x_{n+1}); \nonumber \\
& & [C\psi^{(l,m,n)}](z^\elm, \cz^\emm, x^\enp) = \\
& & =\left. \pa_r \int \psi^{(l,m,n)}(\x_1+r\bo/2, z^\elm; \x_1-r\bo/2 , \cz^\emm; \x_2, \ldots, \x_{n+1}) d^2\om \right |_{r=0}. \nonumber \\
\end{eqnarray}
The proof is in the appendix.

In order to symmetrize $\Si$ let us impose the boundary conditions
\begin{equation}
B \psi= \psi,
\end{equation}
or equivalently
\begin{equation}
\psi^{(l, m, n)}(\x_1, z^\elm; \x_1, \cz^\emm, \x_2, \ldots, \x_{n+1})=
\psi^{(l-1, m-1, n+1)}(z^\elm; \cz^\emm; x^\enp).
\end{equation}
These boundary conditions define the domain $\DD_\Si$. Define finally the operator
\begin{equation}
C_\Si:= C N_e N_{\bar e}
\end{equation}
The symmetric creation/annihilation operator is then
\begin{equation}
\Si_s := \Si + C_\Si.
\end{equation}
The equation
\begin{equation}
\la \phi |\sra{C}_\Si |\psi \ra = \la \phi |C_\Si \psi \ra - \la C_\Si \phi | \psi \ra = \la B \phi |C N_e N_{\bar e}\psi \ra - \la C N_e N_{\bar e} \phi |B \psi \ra
\end{equation}
together with equation (\ref{poti}) proves the symmetry of $\Si_s$.
\section{Discussion}

The main difference between the approach based on the interaction operators (this approach) and the IBC approach is that in this approach the interaction operators can be defined as autonomous and abstract mathematical entities, which make no reference to any physical constants and which can be added to the free Hamiltonian to form the complete Hamiltonian with interaction. This is not the case for the IBC approach.

Another remarkable feature of this approach is the different definition adopted for the emission/absorption operator and for the creation/annihilation operator: in the first case the emitting/absorbing particle is considered as fixed source, and the only Laplacian involved is that of the emitted/absorbed particle. On the contrary, the operator implementing the creation/annihilation process acts in the center of mass of the two particles, and both the particles and their Laplacians are involved on the same footing. While the physical validity of this approach has to be verified with further studies, some formal advantages can already be emphasized.

First of all, in the emission/absorption process particles of different masses may be involved, and therefore an interaction operator acting in the center of mass of the two particles would inevitably include these masses in its mathematical definition. This does not happen for the creation/annihilation operator because the masses of a particle and of its antiparticle are always equal, and therefore the mass does not appear in the coordinates of the center of mass.

Second, we have seen in the non-relativistic domain that the inter-sector interaction only depends on the component of the wave function with null angular momentum, and this is arguably true also in the relativistic domain (see \cite{mo4}). But the center of mass wave function of a pair composed of a fermion (half-integer spin) and a boson (integer spin) cannot have a component with null angular momentum. For this reason an interaction operator acting in the center of mass cannot exist for the emission/absorption of a photon by an electron, as required for example by QED, and the different definition of this operator may allow for a relativistic generalization of the theory.

In conclusion, while this approach seems to have some interesting features, nothing can be said about its validity until some of the results announced for the IBC approach (e.g. rigorous self-adjointness of the Hamiltonian, effective Yukawa potential, correspondence with the renormalization procedure \cite{t2}) are also proved for this approach.

\section*{Appendix} 
%**************************************************************************
\subsection*{Proof relative to the emission/absorption operator} 

In this subsection equations (\ref{inp}) and (\ref{inp2}) are proved. Let us prove first that 
\begin{equation} \label{sse1}
\la \phi^{(l,m)}|\sra{M}|\psi^{(l,m)} \ra = \sum_{k=1}^p \la C_k N_b \phi^{(l,m)} | B_k \psi^{(l,m)} \ra - 
\la B_k \phi^{(l,m,n)} | C_k N_b \psi^{(l,m,n)} \ra.
\end{equation}
To this purpose, let us define the following inter-sector operators from $(l,m,n)$ to $(l, m, n-1)$:
\bea
& & [B^i_k \psi^{(l,m,n)}](y^\ep; x^\en \sm \x_i):=\psi^{(l,m,n)}(y^\ep; \ldots, \x_{i-1}, \y_k, \x_{i+1}, \ldots), \\
& & [C^i_k \psi^{(l,m,n)}](y^\ep; x^\en \sm \x_i):=\\
& & \left.  \pa_r \int_{S^2} \psi^{(l,m,n)}(y^\ep; \ldots, \x_{i-1}, \y_k+ r \bo , \x_{i+1} \ldots) d^2 \om \right |_{r=0}
\eea
By generalizing equation (\ref{a3}) one can easily deduce that 
\begin{equation}
\la \phi^{(l,m,n)}|\sraa{M_k^i}|\phi^{(l,m,n)} \ra = \la C^i_k \phi^{(l,m,n)} | B^i_k \psi^{(l,m,n)} \ra - 
\la B^i_k \phi^{(l,m,n)} | C^i_k \psi^{(l,m,n)} \ra.
\end{equation}
Since the wave function of the bosons is symmetric then $B^i_k=B_k$ and $C^i_k =C_k$. As a consequence:
\bea
& & \la \phi^{(l,m,n)}|\sra{M}|\psi^{(l,m,n)} \ra =\sum_{k=1}^p \sum_{i=1}^n \la \phi^{(l,m,n)}|\sraa{M_k^i}|\psi^{(l,m,n)} \ra = \\
& & \sum_{k=1}^p \sum_{i=1}^n  \la C_k \phi^{(l,m,n)} | B_k \psi^{(l,m,n)} \ra - 
\la B_k \phi^{(l,m,n)} | C_k \psi^{(l,m,n)} \ra= \\
& & \sum_{k=1}^p \la C_k N_b \phi^{(l,m,n)} | B_k \psi^{(l,m,n)} \ra - 
\la B_k \phi^{(l,m,n)} | C_k N_b\psi^{(l,m,n)} \ra= \\
\eea
which of course implies equation (\ref{sse1}). It is then sufficient to prove equation (\ref{inp2}), or simply to prove that 
\begin{equation}
\la \phi^{(l,m,n)}|\sraa{V_{hk}^i}|\psi^{(l,m,n)} \ra = 0,
\end{equation}
For simplicity the equation inpwill be proved for $n=1$, and the following simplifications in the notation will be adopted: (i) the superscript/subscript $1$ will be omitted, so for example $R^1_k$ becomes $R_k$; (ii) the superscript $(l,m,1)$ will be omitted, so for example $\psi^{(l,m,1)}$ becomes $\psi$; (iii) $\De_{\x_1}$ becomes $\De$; (iv) $y^\ep$ becomes $y$.

We have:
\bea
& & \la \phi|\sraa{V_{hk}^1}|\psi \ra \propto \\
& & \la R_h \|\hat \x-\hat \y_h\|^\mo \phi|\De R_k\|\hat \x-\hat \y_k\|^\mo \psi \ra -\la \De R_k\|\hat \x-\hat \y_k\|^\mo \phi| R_h\|\hat \x-\hat \y_h\|^\mo \psi \ra + \\
& & \la R_k\|\hat \x-\hat \y_k\|^\mo \phi| \De R_h\|\hat \x-\hat \y_h\|^\mo \psi \ra -\la \De R_h\|\hat \x-\hat \y_h\|^\mo \phi | R_k\|\hat \x-\hat \y_k\|^\mo \psi \ra = \\
& & \la \phi_h| \sra{\De} |\psi_k \ra + \la \phi_k| \sra{\De} |\psi_h \ra, 
\eea
where
\[
\psi_k:=R_k\|\hat \x- \hat \y_k\|^\mo \psi, \;  \phi_h:=R_h\|\hat \x- \hat \y_h\|^\mo \phi, \tx{ and so on...}
\]
Let us introduce the partial scalar product between two generic vectors $\psi$ and $\phi$ of $\HH^{(l,m,1)}$:
\bg{equation}
\la \phi|\psi \ra_x (y) := \int_{\R^3} \phi^*(y, \x)\psi(y, \x) d^3x, 
\end{equation}
so that
\begin{equation}
\la \phi|\psi \ra = \int \la \phi|\psi \ra_x (y) d^{3p} y
\end{equation}

Let us calculate then $\la \phi_h |\sra{\De} | \psi_k \ra_x (y)$. Let $B_r:= B_r^h \cup B_r^k$, where $B_r^k$ is a 3-ball of radius $r$ centered in $\y_k$, and analogously $B_r^h$. We have:
\bg{eqnarray}
& & \la \phi_h |\sra{\De} | \psi_k \ra_x(y) = \lim_{r \to 0} \int_{\R^3 \sm B_r } \phi^*_h \sra{\De} \psi_k d^3x = 
\lim_{r \to 0} \int_{\pa B_r} \phi^*_h \sra{\na} \psi_k \hat n dS = \nonumber \\
& & \lim_{r \to 0} \left (
 \int_{\pa B^k_r} \phi^*_h \na \psi_k \hat n dS -
 \int_{\pa B^k_r} \na \phi^*_h \psi_k \hat n dS +
 \int_{\pa B^h_r} \phi^*_h \na \psi_k \hat n dS -
 \int_{\pa B^h_r} \na \phi^*_h  \psi_k \hat n dS \right)\; \; \; \; \label{sur}
\end{eqnarray}
where $\hat n$ points inside $B_r$ and the dependence on the variable $y$ is omitted when unnecessary. In the passage from the volume to the surface integral the divergence theorem has been utilized:
\[
\int_{\partial B_r} \phi_h^* \nabla \psi_k \hat n dS = \int_{\R^3 \sm B_r} \nabla (\phi_h^* \nabla \psi_k) d^3x = \int_{\R^3 \sm B_r} \nabla \phi_h^* \nabla \psi_k d^3x + \int_{\R^3 \sm B_r} \phi_h^* \Delta \psi_k d^3x.
\]
By expanding the variable $\x$ of $\psi_k$ and $\phi_h$ in spherical harmonics centered at $\y_k$ the first integral of line (\ref{sur}) becomes:
\bea
& & \ldots = - \lim_{r \to 0} r^2 \int_{S^2} \phi^*_h \pa_r  \psi_k d^2 \om=  - \lim_{r \to 0} r^2 \sum_{\ell m} \phi^*_{h, \ell m}(y, r) \pa_r \psi_{k,\ell m}(y, r) = \\
& & - \lim_{r \to 0} r^2 \phi^*_{h, 00}(y,r) (\pa_r \psi_{00}(y,r)/r - \psi_{00}(y,r)/r^2) = \\
& & = \phi^*_{k, 00}(y,0) \psi_{00}(y,0)= 4 \pi \frac{\phi^*(y,\y_k) \psi(y,\y_k)}{\|\y_k - \y_h\|},
\eea
where the minus sign at the beginning depends on the fact that $\hat n$ points toward the center of the ball. The second integral becomes:
\[
 \ldots = - \lim_{r \to 0} r^2 \int_{S^2} \pa_r \phi^*_h \,  \psi_k d^2 \om=  - \lim_{r \to 0} r^2 \pa_r \phi^*_{h, 00}(y,r) \psi_{00}(y,r)/r = 0.
\]
If the wave functions are expanded in spherical harmonics centered at $\y_h$ the last two integrals of line (\ref{sur}) have the same structure then the first two, and therefore the result is:
\[
\la \phi_h |\sra{\De} | \psi_k \ra_x(y) = 4 \pi \frac{\phi^*(y, \y_k) \psi(y, \y_k) - \phi^*(y, \y_h) \psi(y, \y_h)}{\|\y_h - \y_k\|}.
\]
This expression is antisymmetric in $h$ and $k$, and therefore
\begin{equation}
\la \phi|\sraa{V_{hk}^1}|\psi \ra  = \int \la \phi_h |\sra{\De}|\psi_k \ra_x(y)  + \la \phi_h |\sra{\De} |\psi_k \ra_x(y) \, d^{3p}y =0.
\end{equation}

\subsection*{Proof relative to the creation/annihilation operator}

In this subsection equation (\ref{poti}) is proved. Let us introduce the following inter-sector operators from $(l,m,n)$ to $(l-1, m-1, n+1)$:
\bea
& & [B_{hk}\psi^{(l,m,n)}](z^\elm; \cz^\emm; x^\enp) = \\
& & = \psi^{(l,m,n)}(\ldots, \z_{h-1}, \x_1, \z_h, \ldots; \ldots, \bz_{k-1}, \x_1, \bz_k, \ldots; \x_2, \ldots, \x_{n+1}); \\
& & [C_{hk}\psi^{(l,m,n)}](z^\elm, \cz^\emm, x^\enp) = \\
& & = \left. \pa_r \int \psi^{(l,m,n)}(\ldots, \z_{h-1}, \x_1+r\bo/2, \z_h, \ldots ; \ldots, \bz_{k-1}, \x_1-r\bo/2, \bz_k, \ldots ; \x_2, \ldots, \x_{n+1}) d^2\om \right |_{r=0}.\\
\eea
Let us introduce moreover the following operators acting on the center of mass functions $\psi^\st_{hk}=U_{hk} \psi^\st$:
\bea
& & \Si_\cm:=\frac{R_\z}{\|\z\|} \De_\z \frac{R_\z}{\|\z\|}, \\
& & [B_\cm \psi^\st_{hk}](z^\elm; \cz^\emm; x^\enp) := \psi^\st_{hk}(\ldots, \Z=\x_1, \ldots; \ldots, \z=0, \ldots; \x_2, \ldots, \x_{n+1}),\\
& & [C_\cm\psi^{(l,m,n)}_{hk}](z^\elm, \cz^\emm, x^\enp) = \\
& & = \left. \pa_r \int \psi^{(l,m,n)}_{hk}(\ldots, \Z=\x_1, \ldots; \ldots,  \z=r\bo, \ldots; \x_2, \ldots, \x_{n+1}
 ) d^2\om \right |_{r=0}.\\
\eea
By generalizing equation (\ref{a3}) one can easily deduce that 
\bg{equation} \label{poi1}
\la \phi^\st_{hk} |\sra{\Si}_\cm|\psi^\st_{hk} \ra = \la C_\cm \phi^\st_{hk} |B_\cm \psi^\st_{hk} \ra - \la C_\cm \phi^\st_{hk} |B_\cm \psi^\st_{hk} \ra 
\end{equation}
But one can also easily see that 
\bg{equation} \label{poi2}
\Si_{hk}= U^\mo_{hk} \Si_\cm U_{hk}, \; \; B_{hk} = B_\cm U_{hk}, \; \; \tx{and}\; \;  C_{hk} = C_\cm U_{hk}.
\end{equation}
By inserting the equations (\ref{poi2}) in equation (\ref{poi1}) one obtains
\[
\la \phi^{(l,m,n)} |\sra{\Si}_{hk}| \psi^{(l,m,n)} \ra = \la C_{hk} \phi^{(l,m,n)} |B_{hk} \psi^{(l,m,n)} \ra - 
\la B_{hk} \phi^{(l,m,n)} |C_{hk} \psi^{(l,m,n)} \ra.
\]
Since the fermionic part of the wave function is antisymmetric, one can easily see that $B_{hk}=(-1)^{h+k} B$ and $C_{hk}=(-1)^{h+k}C$. As a consequence:
\bea
& & \la \phi^{(l,m,n)} |\sra{\Si}| \phi^{(l,m,n)} \ra = \sum_{h=1}^l \sum_{k=1}^m \la \phi^{(l,m,n)} |\sra{\Si}_{hk}| \psi^{(l,m,n)} \ra = \\
& & = \sum_{h=1}^l \sum_{k=1}^m \la C_{hk} \phi^{(l,m,n)} |B_{hk} \psi^{(l,m,n)} \ra - 
\la B_{hk} \phi^{(l,m,n)} |C_{hk} \psi^{(l,m,n)} \ra = \\
& & = \left (\la C N_e N_{\bar e} \phi^{(l,m,n)} |B \psi^{(l,m,n)} \ra - 
\la B \phi^{(l,m,n)} |C N_e N_{\bar e} \psi^{(l,m,n)} \ra \right ).
\eea

\bg{thebibliography}{10}

\bibitem{t1} S. Teufel and R. Tumulka: \tit{New type of Hamiltonians without ultraviolet divergence for quantum field theories}, (2015) arXiv:1505.04847

\bibitem{t2} S. Teufel and R. Tumulka: \tit{Avoiding Ultraviolet Divergence by Means of Interior–Boundary Conditions}, in Quantum Mathematical Physics, Springer International Publishing, 293-311 (2016) arXiv:1506.00497

\bibitem{g} B. Galvan: \tit{Regular Hamiltonians for non-relativistic interacting quantum field theories}, (2013) arXiv:1305.0495 

\bibitem{thomas} L.E. Thomas,:\tit{Multiparticle Schr\"odinger Hamiltonians with point interactions}, Phy. Rev. D 30 (1984), 1233-1237.

\bibitem{yafev} D.R. Yafaev: \tit{On a zero-range interaction of a quantum particle with the vacuum}, J. Phys. A:  25 (1992), 963-978.

\bibitem{mo1} M. Moshinsky: \tit{Boundary conditions for the description of nuclear reactions}, Phys. Rev. 81 (1951), 347.

\bibitem{mo2} M. Moshinsky: \tit{Boundary conditions and time-dependent states}, Phy. Rev. 84 (1951), 525.

\bibitem{mo3} M. Moshinsky: \tit{Quantum Mechanics in Fock Space}, Phys. Rev. 84 (1951), 533.

\bibitem{mo4} M. Moshinsky and G. L\'opez Laurrabaquio: \tit{Relativistic interactions by means of boundary conditions: The Breit–Wigner formula}, J. Math. Phys 32 (1991): 3519-3528.

\bibitem{reed} M. Reed and B. Simon: \tit{Method of modern mathematical physics II: Fourier analysis, self-adjointness}, Academic Press, New York (1972)

\end{thebibliography}

\end{document}